# Software Engineering Process Theory:

## A Multi-Method Comparison of Sensemaking-Coevolution-Implementation Theory and Function-Behavior-Structure Theory

Paul Ralph

Lancaster University

paul@paulralph.name

**Abstract**

*Many academics have called for increasing attention to theory in software engineering. Consequently, this paper empirically evaluates two dissimilar software development process theories – one expressing a more traditional, methodical view (FBS) and one expressing an alternative, more improvisational view (SCI). A primarily quantitative survey of more than 1300 software developers is combined with four qualitative case studies to achieve a simultaneously broad and deep empirical evaluation. Case data analysis using a closed-ended, a priori coding scheme based on the two theories strongly supports SCI, as does analysis of questionnaire response distributions (p<0.001; chi-square goodness of fit test). Furthermore, case-questionnaire triangulation found no evidence that support for SCI varied by participants' gender, education, experience, nationality or the size or nature of their projects. This suggests that instead of iteration between weakly-coupled phases (analysis, design, coding, testing), it is more accurate and useful to conceptualize development as ad hoc oscillation between organizing perceptions of the project context (Sensemaking), simultaneously improving mental pictures of the context and design artifact (Coevolution) and constructing, debugging and deploying software artifacts (Implementation).*

# 1. Introduction

The Software Engineering (SE) field has witnessed increasing calls to theorize about its core concepts and processes, e.g. [1]-[8]. However, SE remains preoccupied with normative research on software development methods, methodologies and process models [9] and characterized by "a lack of interest in theories aimed at *understanding* and *explaining* the how and why of the observed design activities" in favor of "a rush from observation and description to prescriptive modelling and the construction of design tools" [10] (p. 153).

Building and empirically evaluating SE theories has many benefits. Theories synthesize, preserve and communicate empirical knowledge, thereby implicitly coordinating future inquiry. Unlike method and tool knowledge, theories endure fashions and fads. Adopting a theoretical mindset furthermore implicitly refocuses researchers on the fundamental rather than superficial features of SE.

A theory is simply a collection of interconnected concepts. Theories have differing purposes including to describe, to explain, to analyze and to predict [11] and units of analysis including individual, group, process, organization and industry [12]. Variance theories focus on *why* events occur while process theories focus on *how* events occur [13]. Variance theories employ different approaches to causation including *regularity* (Y always follows X), *counterfactual* (Y cannot occur without X), *probabilistic* (Y is more likely given X), and *teleological* (X, an agent with free will, chose to do Y) [14]. Similarly, process theories may approximate one of several "ideal types" – *lifecycle* (a sequence of phases), *evolution* (many competing elements), *dialectic* (struggle between several actors with varying power) and *teleological* (goal-oriented, self-directed actions of autonomous actors) [15]. Both types may be used to address a wide variety of questions from *how do developers of aerospace control* *systems formulate unit tests?* to *what are the primary determinants of emotional well-being among video game developers?*

Given the diversity of possible theoretical approaches, deeply understanding sociotechnical phenomena including software development necessitates numerous theoretical perspectives. Following Brooks' [16] insightful elucidation of fundamental confusion surrounding the software development process, this paper focuses on software development process theory. Specifically, it summarizes Sensemaking-Coevolution-Implementation Theory (SCI), which diverges from traditional engineering thinking in an attempt to more accurately explain how software is developed in practice. SCI is evaluated against a rival theory, the Function-Behavior-Structure Framework (FBS), which expresses a more traditional view of the development process. The paper presents an extensive, multi-method, empirical initiative to evaluate these two theories, driven by the following research question.

> ***Research Question:*** *Which of FBS and SCI more accurately represents how teams develop the majority of complex software systems in practice?*

Here a *complex* system is a collection of interconnected elements that exhibits behaviors not predictable from those elements [17]. Complex systems are not necessarily large but exclude routine re-implementation of well-understood artifacts; e.g., a queue data structure. Meanwhile, software development here "encompasses all the activities involved in conceptualizing, framing, implementing, commissioning, and ultimately modifying complex systems" [18] (p. 20). This paper furthermore focuses on 1) development by individuals or coordinated teams predominately working together, rather than projects involving mass-collaboration (e.g., Linux), hostile teams working at cross purposes, or multiple autonomous teams. Additionally, it is primarily concerned with direct actions of development teams, rather than indirect actions and related concepts including project management, politics, power and time.

Section Two discusses process theory in SE, including detailed presentations of FBS and SCI . Section three presents the multi-methodological research design. Section four summarizes the results and section five discusses the study's limitations and implications.

## 2. Related Work

While a comprehensive review of theories used in SE is beyond the scope of this paper, Hannay et al. [4] identified 40 theories that were experimentally evaluated in studies published between 1993 and 2002. However, only two of these were used in more than one article: 1) the Theory of Cognitive Fit, which posits that the alignment between a task and the presentation of information needed for the task affects task performance [19], [20] and 2) the theory that reading techniques affect software inspection effectiveness [21]-[23].

In the following decade, empirical research continued gaining prominence in SE, with, for example, the Empirical Software Engineering and Measurement conference beginning in 2007. However, most empirical work in SE continues either to evaluate specific tools and techniques (e.g., bug prediction approaches [24]) or to investigate specific SE phenomena (e.g., source code clone maintenance [25]). Similarly, most SE theories concern specific SE activities, e.g., search-based testing [26] or visual notation [27]. Meanwhile, little theoretical and empirical work investigates the software development process holistically. Instead, software process research is predominately prescriptive and method-focused [9]. This has produced thousands of software development methods [28] including Scrum [29], Lean [30] and the Unified Process [31], some of which (e.g., the Waterfall Model [32], Spiral Model [33] and Axiomatic Design [34]) are sporadically treated as theories. For example, when Fitzgerald [35] states, “in conventional software development, the development lifecycle in its most generic form comprises four broad phases: planning, analysis, design, and implementation” (p. 589), he is treating Waterfall as a theory.

However, methods are not appropriate foundations for process theories as the former prescribe ostensibly good approaches to an activity while the latter explain the fundamental properties of an activity [10]. Therefore, this sections focuses on process theories, not methods or other prescriptions.

A recent review [36] found no comprehensive software development process theories. However, it did find an engineering design process theory, the Function-Behavior-Structure Framework, which had been applied to SE, and proposed but not empirically test an alternative called Sensemaking-Coevolution-Implementation Theory (discussed next).

### 2.1. Sensemaking-Coevolution-Implementation Theory

SCI (Figure 2; Tables 1 and 2) posits that complex software systems are produced by an agent (individual or team) that alternates between three types of activities in a self-determined sequence. When the design agent is a team, activities may occur in parallel.

Sensemaking, i.e., making sense of an evolving project context, may include interviewing stakeholders, writing notes, organizing notes, reading about the domain, reading about technologies that could be used in the project, sharing insights among team members and acceptance testing (receiving feedback from stakeholders on prototypes). Implementation, i.e., building the software, may include coding, managing the codebase, writing documentation, automated testing, unit testing and debugging.

Coevolution here refers to mutually exploring and refining perceptions of the project context and ideas about existing or potential design artifacts. While Coevolution does not directly map to a variety of well-known software engineering activities, it is observable in real projects. For example, when a team stands around a whiteboard drawing informal models and discussing how to proceed, they often oscillate between ideas about the design object (e.g., ‘how should we distribute features between the partner channel screen and the partner program screen?’) and the context (e.g., ‘you know what, I think channels and programs are just different names for the same thing.’). During Coevolution, ideas about design objects trigger reconceptualization of the project context, which trigger new ideas about design objects, and so on. Coevolution may occur in planning meetings and design meetings, following breakdowns or during an individual’s internal reflection.

Consequently, SCI includes two concentric iterative loops. The inner loop, Coevolution, denotes oscillation between ideas usually over minutes or hours. The outer loop involves making sense of the context, Coevolution and modifying software artifacts, which alter the context and trigger more Sensemaking, usually over weeks or months.

**Figure 1. Sensemaking-Coevolution-Implementation Theory (adapted from [36])**

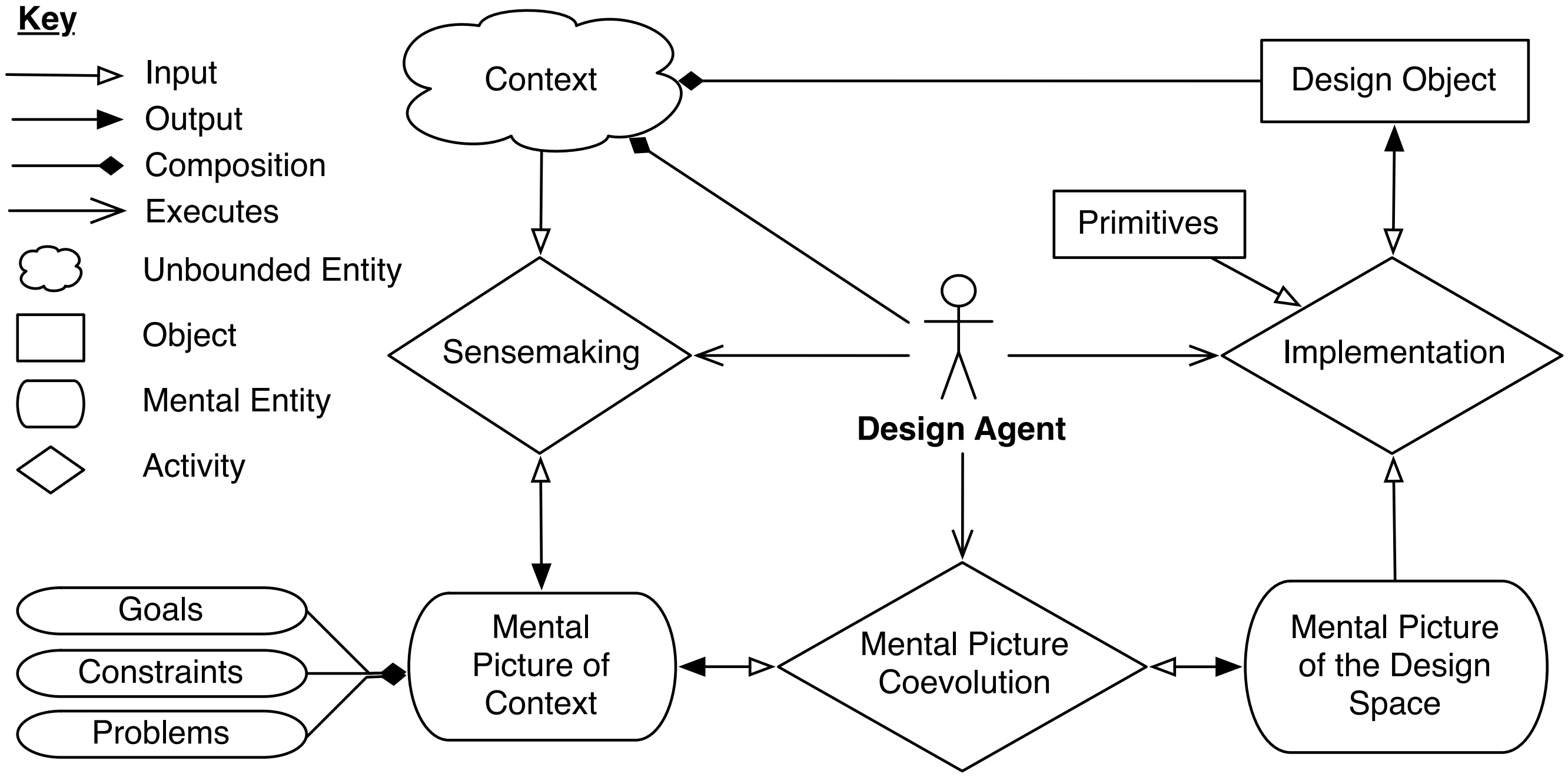

**Table 1. SCI Concepts (adapted from [36])**

| Concept / Activity | Meaning |
|---|---|
| Constraints | the set of all restrictions on the design object's properties |
| Context | the complete surroundings of the design object and agent including the project environment and the object's intended domain |
| Design Agent | an individual or group capable of forming intentions and goals and taking actions to achieve those goals, which specifies the structural properties of the design object |
| Design Object | the thing being designed |
| Goals | the set of optative statements expressed by context actors about the effects the design object should have on its environment |
| Mental Picture of Context | the collection of all of the design agent's beliefs about the context |
| Mental Picture of Design Space | the collection of all of the design agent's beliefs about the design object and its alternatives |
| Primitives | the set of entities from which the design object may be composed |
| Problems | the set of instances of dissatisfaction expressed or experienced by context actors |

**Table 2. SCI Activity Classes (adapted from [36])**

| Concept / Activity | Meaning |
|---|---|
| Sensemaking | the process where the design agent organizes and assigns meaning to its perception of the context, creating and refining the mental picture of context |
| Coevolution | the process where the design agent simultaneously refines its mental picture of the design object, based on its mental picture of context, and the inverse |
| Implementation | the process where the design agent generates or updates the design object using its mental picture of the design object |

## 2.2. The Function-Behavior-Structure Framework

FBS (Figure 2; Tables 3 and 4) posits that engineers design systems by manipulating three kinds of models (i.e., abstract descriptions): function models (F) describe system goals, behavior models describe system requirements ($B_e$) and predicted behavior ($B_S$) and structure models (S) describe system components and their connections. For example, given a project mandate (F), the designer might first *formulate* requirements as a set of use cases (Be) and then *synthesize* a structure as a set of class diagrams. The designer then *analyzes* the structure model (by visual inspection or mathematical simulation) to predict the behavior of the proposed software and *evaluates* the predicted behavior against the use cases. The designer then iterates between synthesis, analysis and evaluation until the structure appears to satisfy the requirements. Sometimes, this process reveals that the requirements or goals are unfeasible, and these are *reformulated*. When the structure appears satisfactory, the designer fleshes out its details and passes on the complete design documentation (D) to programmers.

**Figure 2. The Function-Behavior-Structure Framework (adapted from [37])**

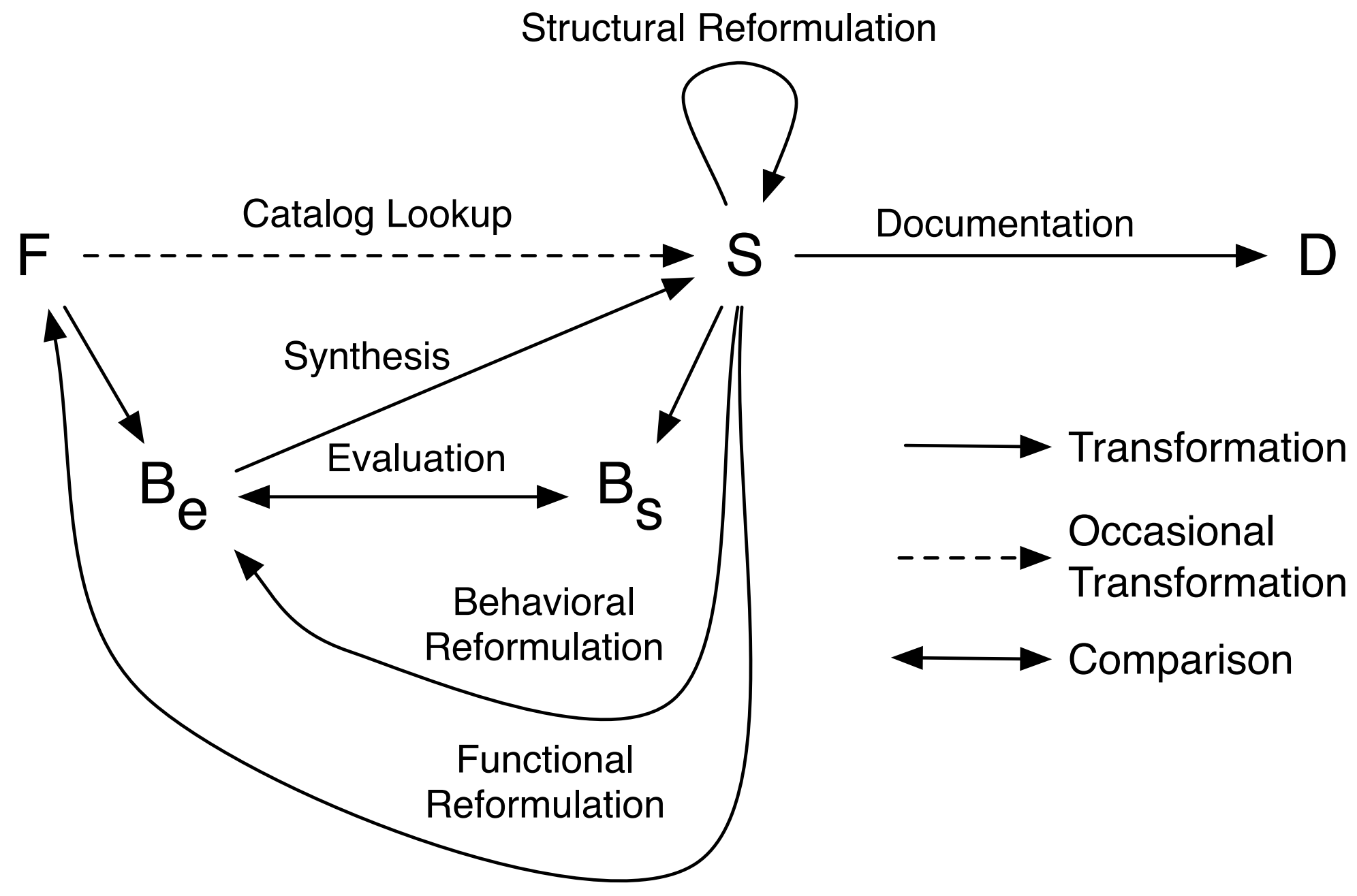

**Table 3. FBS Artifact Classes (adapted from [50])**

| Symbol | Meaning (software) | Definition |
|---|---|---|
| Be | requirement model | expected (desired) behavior of the structure |
| Bs | simulation results | “the predicted behavior of the structure” (p. 3) |
| D | blueprints | a graphical, numerical or textual model that transfers “sufficient information about the designed artifact so that it can be manufactured, fabricated or constructed” (p. 2) |
| F | goal model | “the expectations of the purposes of the resulting artifact” (p. 2) |
| S | design model | “the artifact's elements and their relationships” (p. 2) |

**Table 4. FBS Activity Types (adapted from [50])**

| Activtiy Type | Meaning |
|---|---|
| Formulation | deriving expected (desired) behaviors from the set of functions |
| Synthesis | “expected behavior is used in the selection and combination of structure based on a knowledge of the behaviors produced by that structure” (p. 3) |
| Analysis | the process of deriving the behavior of a structure |
| Evaluation | comparing predicted behavior to expected behavior and determining whether the structure is capable of producing the functions |
| Documentation | transforming the structure into a design description that is suitable for manufacturing |
| Structural Reformulation | modifying the structure based on the structure and its predicted behaviors |
| Behavioral Reformulation | modifying the expected behaviors based on the structure and its predicted behaviors |
| Functional Reformulation | modifying the set of functions based on the structure and its predicted behaviors |

Gero and Kannengiesser [38] extended FBS to elaborate how each model may have different versions in different “worlds”, for example, the goals that the system would ideally achieve vs. the goals it is realistically expected to achieve. While FBS was intended originally to explain engineering design, it has also been applied to software development (cf. [37], [39]).

## 2.3. Conceptual Evaluation of FBS and SCI

FBS and SCI are good rivals for empirical testing for several reasons. First, neither theory has received significant empirical testing.

Second, they make similar kinds of knowledge claims, which significantly differ from the knowledge claims of development methods. SCI and FBS are not simply analogues of Agile and Plan-Driven approaches. Plan-Driven approaches assume that better project planning and more upfront analysis positively impact success. Contrastingly, FBS posits that designers manipulate three types of models and says nothing about project planning or upfront analysis beyond the assumption that systems goals are known or given. While Waterfall, for example, prescribes a series of phases through which a development project progresses, FBS has no phases. Meanwhile, Agile approaches assume that greater success will result from focusing on individuals, software, customer collaboration and responsiveness rather than processes, documentation, contraction negotiation and planning. Contrastingly, SCI makes no predictions concerning success antecedents and simply posits that developers oscillate between three core activities. While Extreme Programming, for example, prescribes rapid analyze-design-build-test cycles [40], SCI suggests that “analysis”, “design”, “coding” and “testing” are fundamentally misleading categories for development activities. One purpose of studying process theory including FBS and SCI is to reveal fundamental concerns obscured by the Agile/Plan-Driven debate.

Third, FBS and SCI are similar enough to test meaningfully. They are both teleological process theories – explanations of how and why an entity changes wherein change is manifested by a goal-seeking agent that engages in activities in a self-determined sequence [15], [41], [42]. They both seek to explain design by organizing activities into several classes (e.g., synthesis, Sensemaking). They both involve models – FBS explicitly and SCI in that designers may externalize their cognition about the context into conceptual models (e.g., user stories) and the design space into design models (e.g., class diagrams). Finally, they both involve iteration.

Fourth, FBS and SCI are different enough to make comparing them interesting. FBS assumes that problem framing, design and implementation are loosely coupled while SCI assumes they are tightly interconnected. Mose specifically, FBS assumes the system goals are given while SCI assumes that designing the system helps to determine its goals. Similarly, FBS assumes that design is model-focused and separate from coding while SCI assumes design is code-focused and models are secondary. Furthermore, FBS posits that the artifact's structure is driven by its requirements, which are driven by goals, while SCI posits that project goals and artifact structure are simultaneously co-created. Moreover, FBS posits that designers primarily evaluate their designs by predicting behavior from design models while SCI posits that designers primarily evaluate their designs by observing the effects of their resulting software artifacts.

More generally, SE research manifests two broad conceptions of design [43]-[46]. One views design as a methodical, plan-centered, approximately rational process of identifying and optimizing a design candidate for known constraints and objectives. In this view, design is loosely coupled with problem framing and implementation. The other views design as an amethodical, improvised, emotional process of simultaneously framing the problem and imagining solutions and constructing artifacts for an unstable, ambiguous context. In this view, design becomes a synonym for development. As SCI and FBS express opposite views, comparing them may provide insight into these conflicting paradigms.

Fifth, both theories bring some *a priori* credibility. SCI originated in the SE field to explaining SE process phenomena [36]. It synthesizes highly influential previous research including Reflection-in-Action [47], Alexander's [48] model of the self-conscious design process, and design coevolution [49]. Meanwhile FBS is itself widely cited, and has spawned a stream of research extending beyond its creator, for example, [10],[37]-[39],[50]-[54]. While FBS may seem prescriptive at times, its developers emphasize that it is predominately explanatory [50] [38]. Moreover, while it could be argued that FBS does not apply directly to software, several papers argue that it does (e.g. [37], [55]). Finally, while FBS may not be the perfect rival theory, methodological guidance strongly suggests using rival theories (below), which are always imperfect, and furthermore no clearly superior alternative was evident.

# 3. Methodology

Process theory testing differs from variance theory testing in several ways. As process theories are concerned with explaining a contemporary phenomenon rather than a causal relationship, they have neither independent nor dependent variables and therefore cannot be tested experimentally. Instead, process theories are best tested using questionnaires or field studies [56], [57], and combining the two increases rigor [58]. Consequently, this section describes a multi-methodological approach combining a multiple-case study to enhance depth and a questionnaire study to enhance breadth within a primarily positivist epistemology. Here “multiple-case study” refers to an empirical inquiry of a contemporary phenomenon that triangulates across multiple locations and data types. The methodology design was informed by commonly used guidelines for questionnaire (e.g. [59]-[61]), positivist case study (e.g. [62], [63]) and multi-method (e.g. [58]) research. The unit of analysis is the team process and all team members are assumed capable of informing on the process.

## 3.1. Hypotheses

Process theory hypotheses are best stated differently from variance theory hypotheses. A causal theory positing that independent variable A causes dependent variable B may be clearly supported by experimentally manipulating A. However, when a process theory is evaluated in the field, at least some observations will support it unless it is absurd and at least some observations will contradict it unless it was overfit to the domain. Consequently, process theories are best evaluated against rival theories [63], [64] – hence the following rival hypotheses.

> ***Hypothesis $H_1$:*** *SCI more accurately reflects how software is created in practice than FBS.*

> ***Hypothesis $H_2$:*** *FBS more accurately reflects how software is created in practice than SCI.*

## 3.2. Instrument development

A case study interview guide (Appendix A) was developed prior to the first case and refined throughout the process. A case study coding scheme (Appendix B) was initially developed such that each concept and relationship in SCI and FBS was given two columns – *evidence for* and *evidence against*. Soliciting feedback on the coding scheme with a colleague familiar with both theories resulted in minor changes.

A pilot (C1, below) was conducted to evaluate the interview guide and coding scheme. After minor improvements both were considered sound and the pilot demonstrated that the relevant phenomena were practically observable. The pilot's results are included in the cross-case analysis as no significant methodological differences from subsequent cases were evident. The pilot also informed questionnaire development, which followed an eight-step process.

1. The author identified differences between the two theories based on their formal descriptions and manifestations in the pilot case.
2. A colleague with expert knowledge of software design reviewed these differences, finding no omissions, biases or unwarranted differences.

3. The author generated approximately 80 items concerning these differences.
4. Items were reviewed by two colleagues with experience in questionnaire-based research, and design practice, respectively.
5. Items were revised and a draft questionnaire was created.
6. A pilot was conducted with three professional developers and seven PhD students to get research-oriented feedback. Items were revised to enhance validity.
7. A second pilot with 12 professional developers was conducted. Items were revised to enhance clarity and brevity.
8. A third pilot with 10 professional developers was conducted. No substantial changes were deemed necessary.

In summary, the pilot case informed both questionnaire and case-study methods, which were used simultaneously to enhance breadth and depth respectively. The final version (Appendix C) comprised 13 items formulated as five-point bipolar scales. Consistent with comparative testing, items examined differences between SCI and FBS rather than specific propositions of either theory. To limit length and increase response rates, the questionnaire focused on three core differences. Each item therefore had one pole indicating agreement with FBS and the other indicating agreement with SCI on one the following differences: 1) whether system goals are given or constructed by the designer (5 items), 2) whether designing and coding are separate or entangled (5 items); 3) whether designing is model-focused or code-focused (3 items). The question order was randomized and the scales were reversed for some questions (e.g., sometimes the SCI pole was on the left, other times on the right). Demographic and project-related questions were also included (see below).

### 3.3. Sampling

The population of interest includes all members (i.e., managers, analysts, etc. – not just programmers) of all software development teams worldwide. For practical purposes, the population was limited to English speakers. As no comprehensive population list was found either globally or for a specific country, random sampling was impractical. The questionnaire was distributed through Twitter, blogs and online social networks including Facebook and LinkedIn to maximize responses through viral invitation. Link tokens were used to record the origin of respondents. Meanwhile, case site selection followed a literal replication strategy [63]. Case studies are nonstatistical, nonsampling research [65]; consequently, the four selected cases are not a representative sample. Their purpose is to explore in-depth manifestations of FBS and SCI elements in practice, not to provide statistically generalizable results. Site selection was constrained by organizational willingness to participate.

### 3.4. Case Context Summary

Four cases were conducted. They vary on several dimensions (Table 5).

**Table 5. Case Diversity**

| Case | Country | Sector | Method | Arrangement | Product | Project size |
|---|---|---|---|---|---|---|
| 1 | Canada | eBusiness | Scrum (agile) | commercial-of-the-shelf | novel product | 5 participants / multi-year |
| 2 | UK | eCommerce | ad hoc (amethodical) | outsourced | novel product | 5 participants / multi-year |
| 3 | UK | education | PRINCE2 (plan-driven) | in-house | legacy system replacement | 25-100 participants / multi-year |
| 4 | UK | education | Scrum-like (agile) | internal entrepreneurship | novel component | 7 participants / 8 months |

Case One (C1) is a mid-sized software services and development company in Vancouver, Canada, which includes several distinct teams. The studied team has five members – two professional web developers, an intern developer, a product owner and a quality assurance analyst. It builds and maintains an online application that helps businesses manage their relationships with their partner organizations. It employed a Scrum-informed process cf. [29]. Originally conceived as a pilot case, the data, analysis and results of C1 were reviewed extensively by a colleague with expertise in case research.

Case Two (C2) is a web development and online marketing agency of between 40 and 50 employees in England. Rather than discrete project teams, the company operates as a hub-and-spokes network where each project is lead by a manager (hub) who assign tasks to whoever has the necessary expertise (spokes) such that each developer's time is split between several simultaneous projects. The case focused on three developers, a graphics designer and the account manager who collaborated on a specific consumer e-commerce website. The project employed an evolving, ad hoc approach.

Case Three (C3) is a mid-sized English university developing and deploying a Moodle-based virtual learning environment. As in C2, participants split their time among many projects. The team was governed by a complicated management structure based on PRINCE2 [66]. In addition to several layers of governance, the project involved three core developers, a technology strategist, a project manager, and minor contributions from dozens of other participants.

Case Four (C4) is a team of part-time developers assembled within a university context to complete a series of small projects, including a mobile application to report facility faults (e.g., broken windows). The team employed a Scrum-like approach and consisted of four developers, a business analyst, a Scrum Master and a Product Owner. Team members were a mixture of BSc and MSc students with industry experience varying from none to three years'.

### 3.5. Data Collection

Between December 2, 2009 and January 11, 2010, 1384 participants from 65 countries responded to the survey. As the sample size is undefined, the response rate cannot be calculated. However, of 4410 individual visitors, 1384 completed the survey (31%), 1118 partially completed it and 1908 bounced. Meanwhile, Cases 1 and 2 involved intensive on-site data collection over periods of two and six weeks respectively, followed by intermittent contact with informants to ask clarifying questions and validate conclusions. C3 involved intermittent data collection over approximately one year. C4 involved intensive data collection 1-2 days per week for eight months, of which five months were spent on the fault reporting project considered here. Although all cases involved semi-structured interviews, the primary mode of data collection for Cases 1, 2 and 4 was direct observation of participants, which produced extensive field notes. Data collection also involved unstructured interviews, copying relevant documents, observing and recording meetings, photographing working conditions and collecting relevant email (Table 6). Each case had its own data collection protocol; all collected data was digitized (if necessary) and held in a single case database.

**Table 6. Data Collection by Case**

| Case | Interviews | Documents | Observation | Meetings | Photos | Emails |
|---|---|---|---|---|---|---|
| 1 | ✔ | ✔ | ✔ | ✔ | ✔ | |
| 2 | ✔ | ✔ | ✔ | | ✔ | ✔ |
| 3 | ✔ | ✔ | | ✔ | | |
| 4 | ✔ | ✔ | ✔ | ✔ | ✔ | ✔ |

### 3.6. Data Analysis

For each case, data analysis began shortly after collection to facilitate adjusting interview questions for unexpected phenomena. Video and audio recordings were transcribed by either the researcher or a professional transcriber. The analysis then proceeded in roughly four phases. First, questionnaire data was statistically analyzed (see below). Second, case study evidence was coded, i.e., the researcher read all transcripts and field notes and copied relevant quotations and excerpts into the predefined coding scheme (above / Appendix B). The same quotation or excerpt could be placed under several categories. This was followed by cross-case analysis and case-questionnaire triangulation. C1 data collection began in April 2008; triangulation completed in April 2013.

# 4. Results

In the interest of space, case-by-case analysis is omitted in favor of cross-case analysis, questionnaire data analysis and case-questionnaire triangulation.

## 4.1. Cross-Case Analysis

Numerous propositions may be derived from FBS and SCI. This section evaluates a selection of core propositions associated with each theory (symbols, e.g. $F \rightarrow B_e$, refer to Figure 2).

*$SCI_1$: Designers engage in Sensemaking.* In SCI, *Sensemaking* includes investigating the context, organizing one's understanding of the context and obtaining feedback on artifacts. All four cases involved investigations, including calling stakeholders with specific questions (C1), face-to-face meetings with clients (C2, C4) and extensive stakeholder consultation (C3). One designer explained "one of the most important parts of my job is to be talking with people on a regular basis, whether they're existing customers or potential customers or just people in the market in general" (C1). In all four cases, designers organized their understanding of the domain, for example, writing call reports summarizing stakeholder comments (C1), writing notes on stakeholder meetings (C2, C3, C4) and researching project infrastructure options (C4). One analyst explained, "I then compiled my notes into much more of a narrative where I've grouped things under themes" (C3). Similarly, all four cases involved feedback on artifacts, e.g., "I would bring the [alpha release] out with me when I'm talking to people and show them" (C1); "Fault report widget delivery meeting ... the facilities and [information systems] guys were reasonably pleased with the interface" (C3 field notes; 24 July 2012); "we get the clients to do some testing" (C2). Therefore, $SCI_1$ is supported.

*$SCI_2$: Designers coevolve their of mental pictures of the context and design space*. In SCI, *Coevolution* specifically refers to oscillating between context understanding and design space understanding where changes to the former trigger changes to the latter and vice versa. Coevolution was directly observed in C1, C2 and C4. For example, the context in C4 was originally framed as helping students report faults in their dorm rooms. Consequently, initial mock-ups did not ask for the location of the fault. This appeared counterintuitive to the developers, and triggered contextual reframing such that students could submit faults anywhere on campus. This triggered design space reframing manifested by adding a location question to the mock-ups. Figure 3 provides a more nuanced example. Coevolution was not directly observed in C3 as developers did not consent to direct observation. Therefore, $SCI_2$ is mostly supported.

**Figure 3. Coevolution in situ**

| |
|---|
| D. draws a diagram of the relationships between the concepts (channel/partner/etc) and the IT artifacts involved, i.e. the online forms, fields and values. They discuss how one would change the site so that we could have different forms, for different kinds of partners. Both hold markers, both edit the diagram, using it to guide their discussion and communicate. This is clearly a reflective, dialectic design process. The diagram mixes the problem and solution, form and context. They seem to be running rapid mental simulations (i.e. what-if analysis). D. draws out the possible combinations of partners/channels/etc. A. checks with M. on the business rules. D. and A. use M. as their "context." D. goes through all possible cases and how they will fit into the proposed design model. More what-if analysis. A. projects into the future: what will the product be like eventually, and how will this effect our design? Their discussion is grounded in their sketching: lots of "this is related to this" with pointing. Very visual. A. proposes new concepts (he calls "classes") to possibly solve the design issue. (C1 Field Notes; 15 Apr 2008) |

$SCI_3$: *Designers implement the design object*. In SCI, *Implementation* includes coding, technical testing and deployment. Coding and technical testing were directly observed in C1, C2 and C4. While not directly observed, coding in C3 was evident from the regular arrival of new code. The nature of technical testing varied substantially. In C2 and C4, testing was quite ad hoc – “We do test ourselves but not really in a systematic way, there is never time for that” (C2). Unit testing was used extensively in C1 and proposed but abandoned in C4 due to the use of a proprietary mobile application framework that impeded unit testing. All four cases employed visual inspection (i.e., black-box testing) wherein developers would observe the effects of their changes, clicking buttons and entering text to see if the site performed as expected. This type of testing was tightly coupled with coding, e.g. “D. makes changes to a copy of the website running on her own development machine. She regularly switches between the code and the website to see what effect her changes have had” (C1 field notes; 16 Apr 2008). Concerning deployment, C1 and C2 had both deployed commercial versions by the end of the data collection period – both are publicly available functional products. C3 and C4 both deployed prototypes to limited user groups during the data collection period and (at the time of writing) have both gone on to release full versions to their entire user bases. $SCI_3$ is therefore supported.

*$FBS_1$: Designers engage in formulation ($F \rightarrow B_e$).* In FBS, *formulation* refers to deriving a behavioral requirements model from a goal model. In cases 1 and 4, no artifact – textual or diagrammatic – approximating a goal model was observed. Case 3 included many documents (at least 200) including a *project mandate*, which ostensibly clarifies the project’s goals and scope. However, the project contained only vacuous goal statements, e.g. “The aim is to provide “sector leading” provision for the 2012 intake of students”. Participants were unable to state meaningful goals, instead making statements including “One of the fundamental things from my point of view is that it must work”. Similarly, participants in case 2 wrote a *project brief*, ostensibly to capture project goals. However, the brief focused on product features rather than goals, as admitted by its writer: “normally I spend a few hours, like 3-4 hours, just looking at the features and the solution". Another participant admitted "I think sometimes we missed the core aim of what the project was trying to achieve". No other documents in cases 2 and 3 contained meaningful goal statements. Therefore, $FBS_1$ was not supported.

*$FBS_2$: Designers engage in synthesis ($B_e \rightarrow S$).* In FBS, *synthesis* refers to devising a structural model of an artifact intended to satisfy a requirements model. In C4, no artifact – textual or diagrammatic – approximating a requirements model was observed. A "requirements" document was observed in C1 and C3 and the "technical specification" in C2 ostensibly served the same purpose. However, at least one C1 team member was unaware of any requirements document and another explained "we think stories instead of requirements". A story, such as "partner application creation is necessary or not for channel creation?" was understood as "a promise to have a conversation" rather than a requirement. Similarly, in C2, although the technical specification was intended to drive the design process, in practice it contained insufficient detail. One participant explained "we don't really write everything down", while another complained "the problem with [the project] was that ... the specification for that was very very brief" and a third admitted that the technical specification "quickly went out of the window because of the volume of changes". Likewise, although C3 included a substantial requirements elicitation and modeling process, it occurred circa spring 2011, while the major design decisions, including using Moodle and its plug-ins, were made in late 2010. Rather than driving design modeling, the "consultation" process was used to justify a priori decisions, leading one participant to malign it as a "pseudo-consultation". Consequently, $FBS_2$ was not supported.

*$FBS_3$: Designers engage in analysis and evaluation ($S \rightarrow B_s \leftrightarrow B_e$).* In FBS, *analysis* refers to predicting the behavior of the proposed structure and *evaluation* refers to comparing the predicted behavior model against the requirements model. Nothing approximating predicting behavior from design models was observed in any of the cases. All four cases included design models, especially website wireframes (C4) and Photoshop renderings / visual mockups (C1, C2, C3). However, nothing like a model of predicted behavior was evident and no observations suggested behavior prediction. Contrastingly in C1, for example, participants evaluated the website (not a design model) by observing (not predicting) its behaviors – "D. makes changes to a copy of the website running on her own development machine. She regularly switches between the code and the website to see what effect her changes have had" (C1 field notes; 16 Apr 2008). $FBS_3$ is therefore not supported.

*$FBS_4$ ($SCI_4$): Problem framing, design and artifact construction are weakly (strongly) coupled*. FBS assumes that design begins with given system goals and ends when detailed design documentation is passed to developers for coding; SCI antithetically posits that problem framing, coding and deployment are all tightly coupled with design. As discussed above, none of the cases exhibited the type of goal models FBS posits. Moreover, no evidence of detailed design documentation was observed – C1 and C4 developers built the software directly from their mental pictures of the system and context, only occasionally referring to user stories, wireframes and mockups; C2 developers worked from an admittedly brief and vague technical specification. Meanwhile, the labyrinth of documents produced by C3 participants may have constituted detailed documentation; however, most of the key design decisions were made *before* the documentation. Instead, participants appear to develop their ideas of project context and design artifact simultaneously (*$SCI_2$* above). Moreover, in C1, C2 and C3, product deployment was an ongoing activity with updates including new features every few weeks and minor fixes and tweaks even more frequently. No separate deployment or transition phase was evident; for example, one participant explained that the "project is more about continuous development and improvement" (C2). In C3, deploying the prototype was a mechanism for understanding the project context – "The VLE pilot phase ... will allow us to learn as much as we can about the issues we will come up against in the larger project" (Project Mandate). In C4, however, the artifact was not deployed during the observation period. Therefore, $SCI_4$ is mostly supported and $FBS_4$ is mostly unsupported.

In summary, none of the four FBS propositions are supported, SCI's Sensemaking and Implementation propositions are strongly supported while its Coevolution and tight coupling propositions are somewhat supported (Table 7). The above analysis concentrates on differences between the two theories. It does not include assumptions they share, including the existence of a design agent and teleological causation, or less central (and less controversial) propositions including SCI's hypothesis that the project context includes constraints and FBS's hypothesis that designers reformulate (i.e., edit) design models.

**Table 7. Support for Selected FBS and SCI propositions**

| | Proposition | Case 1 | Case 2 | Case 3 | Case 4 |
|---|---|---|---|---|---|
| FBS1 | Formulation | ✘ | ✘ | ✘ | ✘ |
| FBS2 | Synthesis | ✘ | ✘ | ✘ | ✘ |
| FBS3 | Analysis/Evaluation | ✘ | ✘ | ✘ | ✘ |
| FBS4 | Loose Coupling | ✘ | ✘ | ✘ | ? |
| SCI1 | Sensemaking | ✔ | ✔ | ✔ | ✔ |
| SCI2 | Coevolution | ✔ | ? | ✘ | ✔ |
| SCI3 | Implementation | ✔ | ✔ | ✔ | ✔ |
| SCI4 | Tight Coupling | ✔ | ✔ | ✔ | ? |

## 4.2. Questionnaire Data Analysis

Respondents varied substantially across demographic, project and company variables (Table 8) but were overwhelmingly male (1241 men vs. 56 women). Most respondents came from the United States (549), Canada (176), the United Kingdom (118) and Australia (73). The most common roles were developer (1325), analyst (569), quality assurance (533), manager (266) and graphics (195). When asked “is your project more ‘social’ (like a website) or ‘technical’ (like a device driver)?”, participants answered more social (34%), more technical (29%) and in between (36%). Based on employers it can be inferred that at least the following sectors are represented: aerospace, applications development, digital media, eCommerce, education, finance, IT consulting, journalism, marketing, networking, operating systems, research, security, sports, telecommunications and tourism.

**Table 8. Summary of Sample Demographics**

| Dimension | Mode or Mean | Minimum | Maximum |
|---|---|---|---|
| Years of Experience | 1 to 5 years (31.5%) | < 1 year (2.9%) | > 25 years (3.6%) |
| Education | Bachelor’s Degree (48%) | some school (1.7%) | PhD (4.1%) |
| Company Size | 1 to 10 (29%) | 1 to 10 (29%) | >10 000 (10.5%) |
| Team Size | 11 members | 1 | 3000 members |
| Project Length | 1.9 years | 0.02 years | 20 years |

Participants responded to 13 items concerning contrasting propositions of SCI and FBS. Assuming responses are coded from 1 (strong support for FBS) to 5 (strong support for SCI), five meaningful results patterns are possible: 1) A positively-skewed distribution (median 1 or 2) indicates that FBS is more accurate; 2) A negatively-skewed distribution (median 4 or 5) indicates that SCI is more accurate; 3) A symmetric distribution (median 3) suggests that SCI and FBS represent extreme positions with most development falling somewhere in between; 4) A bimodal distribution (e.g., modes of 2 and 4) suggests multiple development subcultures, i.e., some developers act as FBS predicts while others act as SCI predicts; 5) A variety of symmetric, positively and negatively skewed items indicates a problem with the survey instrument.

The overall distribution is negatively skewed (Table 9; Figure 4), favoring SCI. The negative skew is significant (p<0.001; $\chi^2$ test) for all items (Appendix D). Although measures of effect size are not available for $\chi^2$, given that 96.6% of respondents had a median response agreeing or strongly agreeing with SCI, the effect size appears large. The practical significance of the observed distribution should be evident from visual inspection of Figure 4. In summary, Hypothesis $H_1$ is supported; $H_2$ is not supported.

**Table 9. Questionnaire Results By Item**

| | | Item | | | | | | | | | | | | | Respondents | |
|---|---|---|---|---|---|---|---|---|---|---|---|---|---|---|---|---|
| | | 1 | 2 | 3 | 4 | 5 | 6 | 7 | 8 | 9 | 10 | 11 | 12 | 13 | Median | Mode |
| Strong FBS | | 7 | 13 | 14 | 20 | 62 | 22 | 22 | 13 | 17 | 58 | 23 | 13 | 9 | 0 | 2 |
| FBS | | 38 | 66 | 42 | 76 | 161 | 61 | 97 | 39 | 63 | 168 | 173 | 174 | 67 | 4 | 21 |
| Neutral | | 72 | 162 | 109 | 120 | 195 | 78 | 113 | 55 | 122 | 148 | 320 | 299 | 303 | 43 | 32 |
| SCI | | 597 | 662 | 576 | 572 | 572 | 398 | 539 | 452 | 539 | 492 | 671 | 623 | 562 | 932 | 717 |
| Strong SCI | | 656 | 446 | 628 | 575 | 349 | 819 | 592 | 796 | 620 | 505 | 173 | 155 | 425 | 406 | 613 |
| **Item** | Median | 4 | 4 | 4 | 4 | 4 | 5 | 4 | 5 | 4 | 4 | 4 | 4 | 4 | | |
| | Mode | 5 | 4 | 5 | 5 | 4 | 5 | 5 | 5 | 5 | 5 | 4 | 4 | 4 | | |

Note: columns do not total 1384 as each question had a "N/A" option

**Figure 4. FBS/SCI Agreement Across 13 Items**

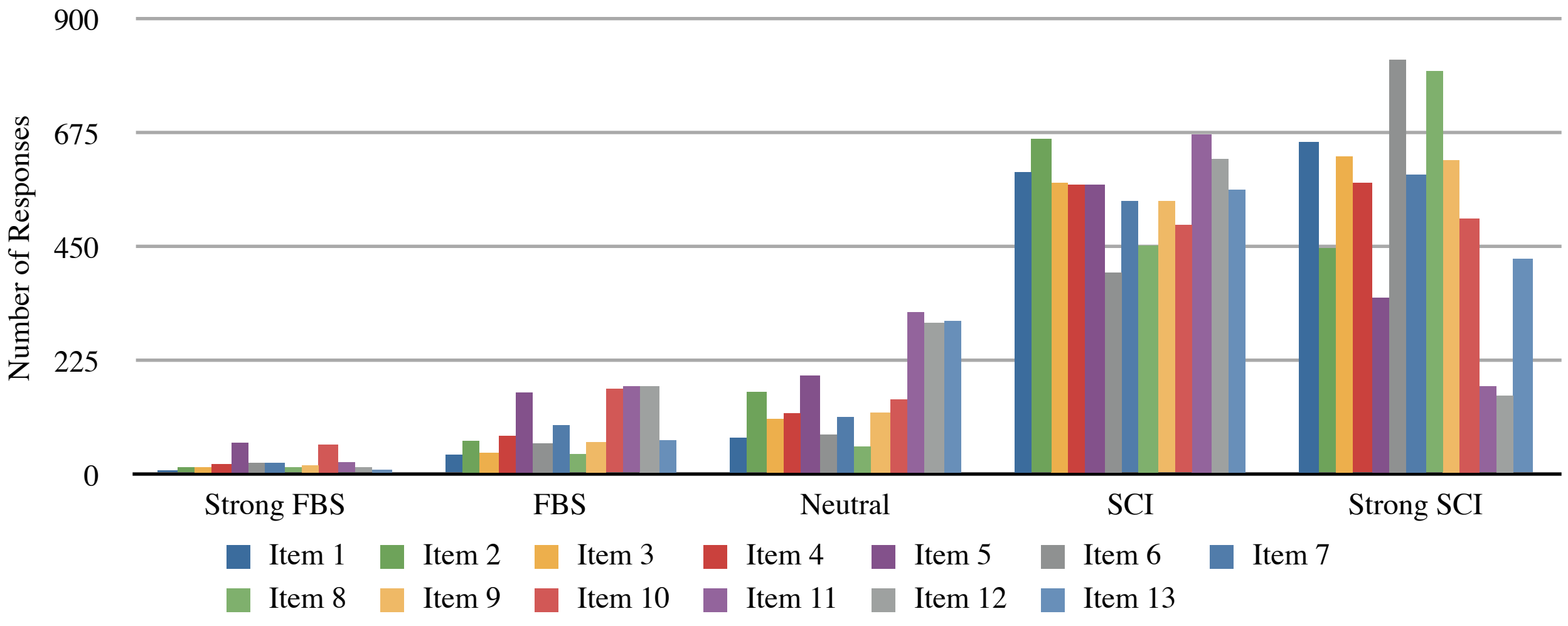

### 4.3. Case-Questionnaire Triangulation

Both questionnaire and case study results favor $H_1$ over $H_2$, i.e., SCI over FBS. Moreover, combining questionnaire and case study results may lead to more nuanced findings.

First, the questionnaire specifically investigates three core differences between SCI and FBS: 1) whether problem framing and design are tightly or loosely coupled; 2) whether design and coding are tightly or loosely coupled; 3) whether respondents primarily work with models or code. Both questionnaire and case data indicate that problem framing, design and coding are tightly coupled. Moreover, case data suggests that deployment of both prototypes and commercial products may be closely interconnected with design in some contexts. While the questionnaire data indicates that respondents principally manipulate source code rather than models, case data indicates that goal models may be absent. Case data further indicates major design decisions may be made independently of requirements models.

Second, combining questionnaire and case studies overcomes many disadvantages of these individual methods. Disadvantages of online questionnaires include inflexibility, limited capacity for open-ended questions and expectancy effects where respondents answer as they believe the researcher expects. In contrast, case studies are very flexible, facilitate open-ended questions and allow the researcher to compare participants' reconstructions of their activities with observations of actual activities. In this study specifically, while the questionnaire was limited to core theoretical conflicts, the case studies facilitated examining individual propositions of each theory.

Meanwhile, questionnaires support statistical generalization while case studies do not [65] – the *n* is too small and representative sampling is impractical. However, as the lack of a software developer population list prevents random sampling for this study, it is reasonable to question whether the survey suffers from sampling bias. Specifically, one could hypothesize that there exists a development subculture (e.g., large firms with formal development processes) for which FBS is more accurate. While such a subculture cannot be definitively established or refuted by this study, some indication of its existence may be found by examining whether responses vary on some demographic or project dimensions.

This requires a measure of an individual's overall orientation toward the frameworks. For the following analyses, an individual's process theory agreement score (or simply "score") is an informal measure of how well a respondent's beliefs correspond to the assumptions of FBS or SCI on a scale of 1 (strong-FBS) to 5 (strong-SCI). This score, defined as follows, is used to simplify presentation of the exploratory analysis and should not be generalized to other meanings or purposes.

> ***Process Theory Agreement Score:*** *a bipolar measure of the extent to which the individual's beliefs about his or her work practices conform to either FBS or SCI, operationalized as the respondent's median response across the 13 items.*

Larger teams were *not* more FBS-like, in fact, participants having median score of 4 or 5 had a *higher* mean team size (11.1 people) than participants having a median score of 1, 2 or 3 (6.7 people); however, this difference was not statistically significant ($p= 0.703$; independent samples t-test). Project length and firm size had no effect on this score. Similarly, median score did not vary by respondent gender, education level, years of experience, country of residence, project role or whether the product was more technical or social in nature. Scores do not vary by self-reported method (Table 10) except that respondents who reported using Lean have a median score of 5 rather than 4. Moreover, scores do not vary by sampling origin, i.e., which advertisement attracted the respondent. In summary, there is no evidence of a contrarian subculture in the survey data.

**Table 10. Median Score by method**

| Method | N | Median Score |
|---|---|---|
| Scrum | 323 | 4 |
| Extreme Programming (XP) | 130 | 4 |
| Agile | 115 | 4 |
| Test Driven Development (TDD) | 89 | 4 |
| Service Oriented Architechture (SOA) | 79 | 4 |
| Other | 62 | 4 |
| Waterfall | 47 | 4 |
| Kanban | 17 | 4 |
| Lean | 16 | 5 |
| Rational Unified Process (RUP) | 14 | 4 |
| “Cowboy Coding” or “Seat-of-the-Pants” | 11 | 4 |

*Notes: 1) Methods mentioned by fewer than ten respondents are grouped as “other”. 2) Some of the “methods” listed by respondents are not technically methods, e.g., SOA is an architectural design pattern. 3) “Median Score” refers to the median of the scores of all respondents indicating the influence of the corresponding method.*

# 5. Discussion and Conclusion

SE theory is crucial to preserve and communicate empirical knowledge and to protect the field against piecemeal empiricism, fads and overemphasis on prescriptive knowledge. This paper consequently examines two dissimilar theories of the software development process. The results suggest that problem framing, problem solving, coding and deployment are tightly coupled activities rather than weakly-coupled phases (as may be inferred from an idealized lifecycle or waterfall model) and that software developers engage in three broad categories of activities – organizing their perceptions of the project context including existing software artifacts (Sensemaking), simultaneously improving their mental pictures of the context and design artifact by oscillating between them (Coevolution), and constructing, debugging and deploying software artifacts (Implementation). Furthermore, while project participants use diverse plans, models and other non-code artifacts, designing is practically entangled with coding.

These results should be interpreted in light of several limitations. First, case data is not statistically generalizable and, as random sampling was impractical, the questionnaire sample may be biased. It is therefore possible that the population includes a more FBS-like subculture. However, given the variety in the reported demographics, suggesting that the entire sample comprises a fringe developer community appears incredulous and no evidence of a subculture was found within the current sample. Second, as the test was primarily comparative, the results do not "prove SCI"; the results simply favor SCI over FBS. Another interpretation of the results is simply that it is more common for software projects to meet SCI's assumptions than FBS's. Clearly, these results say nothing about whether attempting to follow a more FBS-like process would be beneficial. Third, many observed phenomena are not obviously covered by either theory, including the use of informal models, quality, success, management and politics.

With these caveats in mind, the results of this study have numerous implications for educators, practitioners and researchers. For educators, recognizing the centrality of Coevolution in software development motivates major shifts in software engineering curricula, which largely ignores Coevolution [67]. Programs should cover SCI instead of or in addition to the Waterfall Model as the basic form of development and cover Coevolutionary thinking including creativity techniques [68] and sketching [69].

Furthermore, as analysis, design and coding are synchronous and tightly-coupled, attempting temporal separation of these activities as artificial phases or assigning them to separate teams or individuals is likely counterproductive. For example, if teams build their understanding of a system's goals by building the system and obtaining feedback, assigning 'goal analysis' to a 'business analyst' during the 'analysis phase' simply does not make sense. Additionally, as problem framing and solving are simultaneous, interconnected activities, expecting project participants to accurately estimate the time, budget or effort prior to development is simply unrealistic, which may explain the prevalence of inaccurate effort estimation [70]. This suggests that fixed-price/schedule contracts will increase overall project risk [16].

For researchers, these results illuminate fundamental questions often obscured in methods literature. For example, what is the role of requirements engineering if design is driven by Coevolution rather than requirements specification? What are the best kinds of artifacts for facilitating development? What is the role of project management if driving projects through budgets and schedules is unwise? Moreover, this analysis suggests that asking when to use Lean or Scrum or the Unified Process or Waterfall may be the wrong question as actual processes may systematically differ from any of specific method. As the post-methodology era [71] in which developers reject methods in principle solidifies, research may shift focus from methods to individual practices, and to psychology- or sociology-informed antecedents of success, including motivation [72] and cognitive bias [73]. Furthermore, much extension and further analysis of SCI is possible, including exploring the role of non-software artifacts, relaxing the single-agent assumption and clarifying SCI's relationship to different forms of testing.

In conclusion, this study presents the most comprehensive, if not the first, empirical analysis of either FBS or SCI in the domain of software development. Its core contribution is the finding that SCI provides the more accurate account of how most complex software is developed in practice. This conclusion rests on the responses of more than 1300 programmers, analysts, testers and managers from over 60 countries and approximately two years of field research including hundreds of hours of interviews and direct observation. Finally, this paper is intended to motivate greater attention to process theory in SE research and to fundamental assumptions of existing SE paradigms.

# Appendix A: Example Interview Questions

The following is an example of the kind of interview guide used for the case studies. The precise questions asked were tailored to each case, and each participant; however, these questions provide a sense of the core topics. Follow-ups and probes are not included.

1. Introduction
   1. Participant name
   2. Company name
2. Positioning the individual
   1. What is your position at [ company name ] called?
   2. What project(s) are you involved with?
   3. What are you building?
   4. What are your responsibilities?
   5. What are your roles in this project?
3. Theme one: development organization
   1. How is software development organized in your company?
   2. What are the important roles?
   3. Who is involved?
   4. What are their respective responsibilities?
   5. What tools do you use?
   6. What kind of documents do you use?
   7. What kind of models or diagrams do you use?
   8. Who are your clients, customers or users?
4. Theme two: individual activities
   1. What do you spend most of your time doing?

2. On the first day of the last [iteration/cycle/Sprint/whatever is indicated in theme one], what is the first thing you did?
    3. The second?
    4. Then what?
    5. Etc.
5. Conclusion
    1. Is there anything else you want to tell me?
    2. Is there anything else you think I should know?

# Appendix B: Coding Scheme

Table 11 illustrates how qualitative evidence was categorized and organized. The evidence was never actually combined into a single table as it would be impractically large. For example, the notes for Case 1/for/Coevolution comprise 1230 words.

# Appendix C: Questionnaire Items

Although the actual questionnaire was conducted online, its content is more clear in the paper version (below) than it would be from screenshots.

**Table 11. Coding Scheme**

| Concept | Case 1 | | Case 2 | | Case 3 | | Case 4 | |
|---|---|---|---|---|---|---|---|---|
| | for | against | for | against | for | against | for | against |
| **FBS** | | | | | | | | |
| Functions | | | | | | | | |
| Expected Behavior | | | | | | | | |
| Predicted Behavior | | | | | | | | |
| Structure | | | | | | | | |
| Formulation | | | | | | | | |
| Synthesis | | | | | | | | |
| Analysis | | | | | | | | |
| Evaluation | | | | | | | | |
| Reformulation | | | | | | | | |
| Design Description | | | | | | | | |
| Tight Coupling | | | | | | | | |
| **SCI** | | | | | | | | |
| Sensemaking | | | | | | | | |
| Coevolution | | | | | | | | |
| Implementation | | | | | | | | |
| Context | | | | | | | | |
| Mental Picture of Context | | | | | | | | |
| Goals | | | | | | | | |
| Mental Picture of Design Space | | | | | | | | |
| Design Object | | | | | | | | |
| Primitives | | | | | | | | |
| Design Agent | | | | | | | | |
| Loose Coupling | | | | | | | | |

# About You and Your Project

Please answer all questions to the best of your knowledge. If you don't know the answer to a question, skip it, but please do not skip questions unnecessarily. The survey should take 10 to 15 minutes.

1. What is your gender?
   - ◯ Male
   - ◯ Female
   - ◯ Other

2. What is the highest level of education you have attained?
   - Some school
   - High school Diploma
   - Some college, university, trade school, etc.
   - Diploma from technical college, trade school, etc.
   - Bachelor's degree
   - Master's degree
   - PhD

3. How long have you been involved in the software development industry? (In-house, off-the-shelf, for-client, etc.)
   - less than 1 year
   - 1 to 5 years
   - 6 to 10 years
   - 11 to 15 years
   - 16 to 20 years
   - 21 to 25 years
   - more than 25 years

4. What is your current occupation (check all that apply):
   - ☐ business or requirements analyst
   - ☐ graphics designer / animator
   - ☐ team lead
   - ☐ project manager
   - ☐ programmer / developer
   - ☐ quality assurance / tester
   - ☐ other ______________________

5. How many employees does your company have? (Required)
   - 1 - 10
   - 11 - 100
   - 101 - 1000
   - 1001 - 10 000
   - More than 10 000
   - Not applicable / Don't Know

6. Definition: For the purposes of this survey, your project is the software development project on which you are currently working. If you are not currently working on a project, your project is the last one you worked on. If you are working on multiple projects, your project is the one on which you spend the most time. Please answer all questions based on the same project.

What roles have you held in your project? (check all that apply) (Required)

- ☐ business or requirements analyst
- ☐ graphics designer / animator
- ☐ project manager
- ☐ programmer / developer
- ☐ quality assurance / tester
- ☐ other - please list: ________

7. Definition: For the purposes of this survey, your team consists of everyone with whom you collaborate closely on your project, regardless of their geographic location or official title. Your team is all the people you feel like you work with.

Approximately how many people are currently on your development team? (If the project is complete, what was the largest number of people who were on the team at one time?)

________

8. Please list any development methods your team is explicitly using (e.g., RUP, Extreme Programming, SCRUM, Service Oriented Architecture).

________

9. Approximately how long has your project been in progress?

years: ________

months: ________

10. Is your project more "social" (like a website) or "technical" (like a device driver)?

- ○ More Social
- ○ Somewhere in between
- ○ More Technical
- ○ Don't know

# Propositions 1 and 3

## 11. Please indicate the extent to which you agree with the following statements. (Required)

| | Strongly disagree | Disagree | Neutral | Agree | Strongly agree | Not applicable / don't know |
|---|---|---|---|---|---|---|
| No one thing drives all design decisions – they are made based on a variety of information | | | | | | |
| Changes to my team's understanding of what the software is supposed to do were triggered by changes in our understanding of the problem/situation | | | | | | |
| My understanding of what the software is supposed to do has been influenced by several factors (e.g., management, marketing, clients, the dev team, standards, my own values, experience on previous products, etc.) | | | | | | |
| My understanding of the software's purpose has been influenced by several factors (e.g., management, marketing, clients, the dev team, standards, my own values, experience on previous products, etc.) | | | | | | |
| The process of designing the software has NOT helped my team better understand the context in which the software is intended to be used | | | | | | |
| A complete, correct specification of low-level design decisions was available before coding began (*e.g., whether to use a hashtable or array to store usernames) | | | | | | |
| The software was coded iteratively | | | | | | |
| My team has revised the software code based on new information (e.g., bug reports, failed unit tests, feedback from Quality Assurance, etc.) | | | | | | |
| My team now understands what the software is supposed to do better than we did when we started coding | | | | | | |
| Low-level design decisions* were primarily made before the first line of code was written (*e.g., whether to use a hashtable or array to store usernames) | | | | | | |

# Proposition 2

12. Definition: For the purposes of this survey, a model is an abstraction of the software. Models can be diagrammatic (e.g., UML class diagrams, data flow diagrams) or text (e.g., a list of requirements, a set of user stories). Just to be clear, code refers to the source code of the software.

We have found that some developers figure out the detailed design of a product through making and revising models, whereas others do their detailed design by writing and revising code. (Reminder: detailed design refers to things like deciding whether to use a hash table or an array to store passwords).

Please indicate where you and your team do your detailed design work. (Required)

| | Exclusively with models | Mostly with models | Equally with models and code | Mostly with code | Exclusively with code | Not applicable / don't know |
|---|---|---|---|---|---|---|
| I do detailed design... | ○ | ○ | ○ | ○ | ○ | ○ |
| My team does detailed design... | ○ | ○ | ○ | ○ | ○ | ○ |

13. It is possible to test software in two ways:

1) Prediction: testers inspect models of the software and predict how code based on those models will behave (e.g., predict from a UML class diagram how the code will handle an error).

2) Observation: testers run the code and see what it does (e.g., unit testing, manually test the interface).

Which of these is more consistent with how your team does testing? (Required)

- ○ Exclusively prediction
- ○ Mostly prediction
- ○ Both prediction and observation
- ○ Mostly observation
- ○ Exclusively observation
- ○ Not applicable / don't know

# Finishing comments and interview signup

14. Is there anything else you would like to add to your response? Anything you feel we should know?

15. If you would be willing to discuss your results further, please enter your contact info below. (Don't worry, we hate spam as much as you do.)

Name:

Phone # or Email Address:

Employer:

# Appendix D: Expanded Methodology

Many methodologists and statisticians disagree on whether Likert and semantic differential scales produce interval or ordinal data and, consequently, on whether to apply parametric or nonparametric tests[1]. In the interest of caution, this paper uses nonparametric tests. Nonparametric tests (including chi-square) may be used to evaluate the statistical significance of this distribution; however, these tests require an expected distribution to compare with the observed distribution. As no *a priori,* theoretically-justified distribution is available, the "expected distribution" must be generated somehow. Three alternatives are apparent.

1. Uniform Distribution - on any given item, responses split evenly between all categories.
2. Pseudo-Normal Distribution - an approximated normal distribution on a five point scale.
3. FBS-supporting distribution - a distribution that favors FBS to the same extent that the observed distribution favors SCI.

The uniform and normal distributions are automatically generated from the data by the SPSS Statistics software package when using the Kolmogorov-Smirnov (K-S) Test. The FBS-supporting distribution was generated by inverting the observed distribution (i.e., subtracting each response from six).

Using the normal distribution addresses the question *is the extent of negative skew in the observed distribution significant?* Using the FBS-supporting distribution addresses the question *is the observed distribution significantly different from an equally compelling distribution supporting the alternative hypothesis?*

---

[1] Harwell, M. R., & Gatti, G. G. (2001). Rescaling Ordinal Data to Interval Data in Educational Research. Review of Educational Research, 71(1), 105–131.

However, generating a pseudo-normal distribution using the K-S test involves calculations (e.g., mean) inconsistent with interpreting Likert scales as ordinal data; therefore, these statistics should be interpreted with caution. The most defensible test is the chi-square goodness of fit test (with significance via the sign test) using the reflected (FBS-supporting) distributions as this directly compares Hypotheses $H_1$ and $H_2$ with a minimum of assumptions. Given the sample size and magnitude of skewness of the observed distribution, these results are fairly robust against minor changes in the expected distribution. Table 12 shows the results of testing the observed distributions against the three alternative theoretical distributions.

**Table 12. Chi-Square Test Results**

| | Uniform Distribution | | Pseudo-Normal Distribution | | FBS-Supporting Distribution | |
|---|---|---|---|---|---|---|
| **Item** | **K-S Test Z** | **Significance** | **K-S Test Z** | **Significance** | **Sign Test Z** | **Significance** |
| 1 | 24.59 | p < 0.001 | 10.45 | p < 0.001 | -33.49 | p < 0.001 |
| 2 | 21.00 | p < 0.001 | 10.35 | p < 0.001 | -29.9 | p < 0.001 |
| 3 | 23.21 | p < 0.001 | 9.851 | p < 0.001 | -32.21 | p < 0.001 |
| 4 | 21.90 | p < 0.001 | 9.727 | p < 0.001 | -29.92 | p < 0.001 |
| 5 | 15.92 | p < 0.001 | 10.25 | p < 0.001 | -20.47 | p < 0.001 |
| 6 | 23.51 | p < 0.001 | 12.64 | p < 0.001 | -31.45 | p < 0.001 |
| 7 | 21.32 | p < 0.001 | 9.648 | p < 0.001 | -28.53 | p < 0.001 |
| 8 | 24.72 | p < 0.001 | 12.57 | p < 0.001 | -33.18 | p < 0.001 |
| 9 | 22.12 | p < 0.001 | 9.677 | p < 0.001 | -30.48 | p < 0.001 |
| 10 | 17.70 | p < 0.001 | 9.837 | p < 0.001 | -22.13 | p < 0.001 |
| 11 | 13.65 | p < 0.001 | 10.82 | p < 0.001 | -20.1 | p < 0.001 |
| 12 | 12.94 | p < 0.001 | 10.35 | p < 0.001 | -18.84 | p < 0.001 |
| 13 | 17.50 | p < 0.001 | 8.782 | p < 0.001 | -27.87 | p < 0.001 |